\documentclass[12pt,twocolumn]{article}

\usepackage{amsmath,amssymb}
\usepackage{times}
\usepackage{xcolor}
\usepackage{graphicx}
\usepackage[font=small]{caption}
\usepackage{float}
\usepackage[switch]{lineno}
\usepackage[
  top=2cm,
  bottom=2cm,
  left=2cm,
  right=2cm
]{geometry}
\usepackage[
  backend=biber,
  style=numeric-comp,
  sorting=none
]{biblatex}
\DeclareBibliographyDriver{article}{%
  \usebibmacro{bibindex}%
  \usebibmacro{begentry}%
  \usebibmacro{author/editor+others/translator+others}%
  \setunit{\labelnamepunct}\newblock
  \usebibmacro{title}%
  \newunit
  \usebibmacro{journal+issuetitle}%
  \newunit
  \usebibmacro{doi+eprint+url}%
  \newunit\newblock
  \usebibmacro{addendum+pubstate}%
  \setunit{\bibpagerefpunct}\newblock
  \usebibmacro{pageref}%
  \usebibmacro{finentry}%
}
\renewbibmacro{in:}{%
  \ifentrytype{article}{}{\printtext{\bibstring{in}\intitlepunct}}%
}
\renewbibmacro*{journal+issuetitle}{%
  \usebibmacro{journal}%
  \setunit*{\addspace}%
  \iffieldundef{series}
    {}
    {\newunit
     \printfield{series}%
     \setunit{\addspace}}%
  \printfield{volume}%
  \setunit{\addcolon\space}%
  \printfield[bold]{number}
  \setunit{\addcomma\space}%
  \printfield{pages}%
  \setunit{\addcomma\space}%
  \iffieldundef{year}
    {}
    {\printtext[parens]{\printdate}}
  \newunit}
\renewbibmacro*{finentry}{\relax}
\DeclareFieldFormat
  [article,inbook,incollection,inproceedings,patent,thesis,unpublished]
  {title}{#1}

\newcommand{\vbg}{V_\mathrm{{BG}}}

\newcommand{\bfA}{{\textbf{a}}}
\newcommand{\bfB}{{\textbf{b}}}
\newcommand{\bfC}{{\textbf{c}}}
\newcommand{\bfD}{{\textbf{d}}}
\newcommand{\bfE}{{\textbf{e}}}

\newcommand{\W}{{WSe\(_2\)}}
\newcommand{\Mo}{{MoSe\(_2\)}}

\begin{document}

\title{Dipolar excitonic quantum wires at \\ atomically sharp lateral interfaces}

\author{Elie Vandoolaeghe$^{1\dagger}$, Francesco Fortuna$^{1\dagger}$, Suman Kumar Chakraborty$^{2}$, \\ Biswajeet Nayak$^{2}$,  Takashi Taniguchi$^{3}$, Kenji Watanabe$^{3}$, Prasana K. Sahoo$^{2*}$, \\ Thibault Chervy$^{4}$, and  Puneet A. Murthy$^{1\ast}$}

\twocolumn[
\maketitle

\normalsize{$^\dagger$ These authors contributed equally to the work}\\
\noindent \normalsize{$^{1}$Institute for Quantum Electronics, ETH Z\"urich, CH-8093 Z\"urich, Switzerland}\\
\normalsize{$^{2}$Quantum Materials and Device Research Lab, Materials Research Center, Indian Institute of Technology, Kharagpur, India}\\
\normalsize{$^{3}$National Institute for Materials Science, Namiki 1-1, Tsukuba, 305-0044, Ibaraki, Japan}\\
\normalsize{$^{4}$NTT Research, Inc. Physics \& Informatics Laboratories, 940 Stewart Dr, Sunnyvale, CA 94085}\\

\begin{center}
\normalsize{$^\ast$ Corresponding author:\\
Puneet A. Murthy, murthyp@ethz.ch; Prasana K. Sahoo, prasana@matsc.iitkgp.ac.in}
\end{center}
]

\textbf{One-dimensional (1D) quantum systems are a cornerstone of many-body physics. However, their realization in solids has traditionally relied on top-down methods \cite{Giamarchi2003}, which are limited by structural disorder and coarse confinement. Here, we demonstrate a fundamentally distinct route: the emergence of 1D quantum matter at the atomically sharp interface between monolayer semiconductors. Using lateral $\text{MoSe}_2-\text{WSe}_2$ heterostructures, we identify interfacial excitonic quasiparticles that are bound to the crystal junction. Photoluminescence spectroscopy resolves these excitons into a ladder of discrete states, establishing nanoscopic 1D confinement at length scales $\lesssim$3 nm. These excitons possess exceptional large permanent in-plane electric dipole moments exceeding $e \times 2\,$nm, and exhibit micron-scale, highly anisotropic diffusion confined to the interface. Crucially, the lateral geometry enables dynamic, \emph{in-situ} reconfiguration of the exciton’s internal structure. By introducing electrostatic doping, we demonstrate a collapse of the dipole moment and a 20-fold reduction in radiative lifetime. This structural tunability establishes lateral interfaces as a uniquely powerful platform for the `bottom-up' engineering of 1D quantum matter. By enabling the dynamic tuning of wavefunctions within a single atomic monolayer, this work opens a scalable route toward 1D excitonic circuits and strongly correlated 1D bosonic phases.}

\begin{figure*}[!t]
\centering
\includegraphics[width=\linewidth]{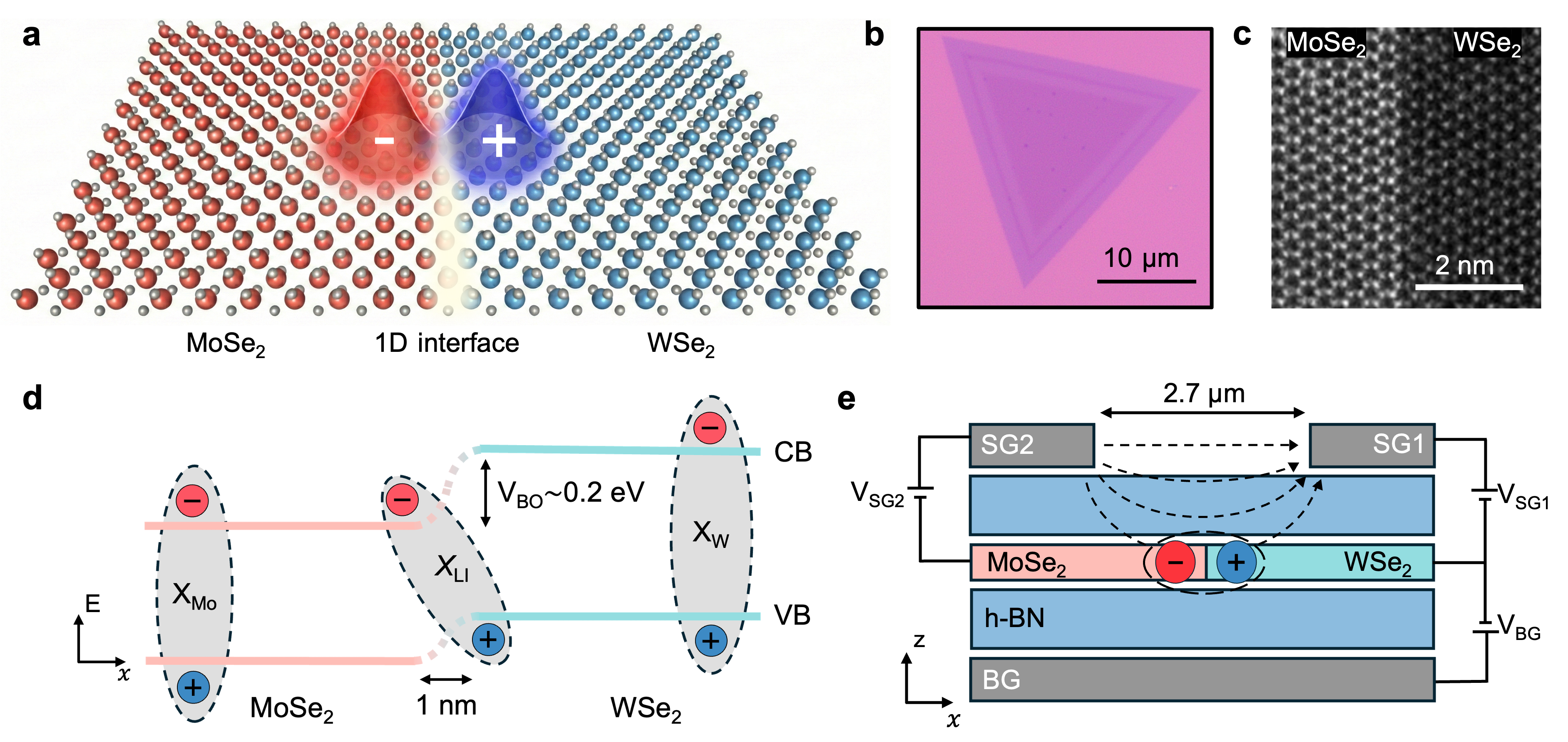}

\caption{\textbf{Lateral interfaces and the interfacial dipolar exciton ($X_{\mathrm{LI}}$)}. 
\textbf{a}, Illustration of the 1D lateral interface between 2D semiconductors \Mo\ and \W, hosting interfacial excitons.
\textbf{b}, Micrograph of CVD-grown monolayer lateral heterostructures (LHS) of \Mo\ and \W. 
\textbf{c}, High Angle Angular Dark-Field Scanning Transmission Electron Microscopy (HAADF-STEM) image confirming the atomic sharpness of the interface. 
\textbf{d}, Band edge diagram illustrating the Type-II alignment, confining the electron in \Mo\ and the hole in \W. $X_{\mathrm{Mo}}$ and $X_{\mathrm{W}}$ denote the bulk direct excitons. The predicted state $X_{\mathrm{LI}}$ is the interfacial dipolar exciton, where the separated charges remain Coulomb-bound at the junction. 
\textbf{e}, Schematic of the device architecture: h-BN (top and bottom are 25 nm thick) encapsulated monolayer LHS, featuring bottom (BG) and split top gates (SG1, SG2) for control over carrier density and in-plane electric field. }
\label{fig:one} 
\end{figure*}

The realization of one-dimensional (1D) quantum systems remains a central challenge in condensed matter physics. In the 1D limit, reduced dimensionality enforces strong particle correlations, giving rise to exotic phases such as Luttinger liquids and Tonks-Girardeau gases \cite{Cazalilla2011,Richard2022}. Furthermore, the restricted kinematic phase space in 1D leads to suppressed scattering from phonons and disorder, potentially enabling high-mobility transport unattainable in higher dimensions \cite{Sakaki1980}. Traditionally, these states are pursued by `carving' 1D channels using top-down lithography or electrostatic gating \cite{Wees1988,Kane1998,Colombo2008,Li2025}. However, these methods are limited by extrinsic disorder, fabrication constraints and confinement scales.

We demonstrate a fundamentally distinct paradigm: the intrinsic emergence of 1D quantum matter at the atomically sharp interface between dissimilar transition-metal dichalcogenide (TMD) monolayers, $\text{MoSe}_2$ and $\text{WSe}_2$ (Fig.\,\ref{fig:one}\,\bfA). Van der Waals (vdW) semiconductors are uniquely suited for this approach, providing boundaries defined with near-atomic precision. Recent advances in chemical vapor deposition (CVD) enable the synthesis of lateral heterostructures (LHS) where two monolayers are seamlessly stitched within a single atomic plane \cite{Huang2014,Duan2014,Li2015,Sahoo2018}. A micrograph of a typical LHS flake is shown in Fig.\,\ref{fig:one}\,\bfB. These junctions form atomically sharp interfaces only a few lattice constants wide yet extending over tens of microns, realizing clean, crystallographically defined 1D boundaries between 2D electronic phases. The atomic sharpness is confirmed via HAADF-STEM imaging of LHS flakes, shown in Fig.\,\ref{fig:one}\,\bfC.

The elementary optical excitations in these systems are excitons —Coulomb-bound electron-hole pairs. In monolayer TMDs, these quasiparticles possess exceptionally high binding energies and their 2D physics is well understood \cite{Wang2018}. While prior works have demonstrated 1D confinement in various lithographically defined or strain-induced nanostructures \cite{Thureja2022,Dirnberger2021}, these typically remain in the `weak' confinement regime, where the confinement length scale ($\ell \gtrsim 10\text{ nm}$) far exceeds the intrinsic exciton Bohr radius ($a_B \approx 1\text{ nm}$). The introduction of a lateral interface, however, introduces a new energy scale from the Type-II band alignment (Fig.\,\ref{fig:one}\,\bfD) \cite{Degani1990}. This band offset partitions electrons and holes into separate regions, directly competing with the Coulomb attraction.

When this band offset is comparable to the Coulomb energy, theory predicts \cite{Kang2013,WangYao2018,Wilson2017,Urbaszek2023,lima2023,DUrnevSmirnov2025} the emergence of a new ground state energetically below the 2D exciton of both parent materials. In this unique charge-transfer state, the electron and hole reside on adjacent sides of the junction while remaining Coulomb-bound. These interfacial excitons ($X_{\text{LI}}$) feature reduced binding energies but exceptionally large, permanent in-plane dipole moments. Crucially, while the carriers are spatially separated, the exciton's center-of-mass is strictly bound to the 1D junction. Unlike vertical vdW heterostructures \cite{Rivera2015,Robertson2016,Jiang2021}—where the dipole is geometrically ``frozen" by the layer spacing—this 2D architecture enables the dipole to exist within the junction plane, providing a uniquely tunable platform for the in-situ engineering of 1D quantum matter.

Despite intense theoretical and experimental interest, the realization of 1D quantum confined dipolar interfacial excitons has been elusive. While prior works have reported preliminary evidence for charge transfer (CT) excitons \cite{Urbaszek2023} as well as exciton transport in LHS \cite{Yuan2023, Beret2022,Sahoo2024,Rosati2025}, they have been limited by interface width and disorder, leading to broad and weak spectral features \cite{Chu2018, Sahoo2024}. Consequently, the ability to resolve the discrete quantum structure and actively engineer the radiative dynamics of these states has remained an outstanding challenge.

In this work, we realize the regime of strongly quantum-confined 1D dipolar excitons by bridging atomic-scale structural precision with active electrostatic control. 
To preserve the near-atomic sharpness of the junctions and reduce effects of ambient dielectric disorder, the LHS monolayer flakes are encapsulated in hexagonal boron nitride and integrated into a dual-gate architecture (Fig.\,\ref{fig:one}\,\bfE). This configuration—featuring a global back gate (BG) and split top gates (SG$_1$, SG$_2$)—enables the independent regulation of carrier density and in-plane electric fields, a critical capability for probing the exciton’s internal structure. By combining high-resolution spatio-temporal spectroscopy with this multi-gate control, we provide an unambiguous demonstration of both the fundamental 1D nature and the functional tunability of these interfacial states.

\begin{figure*}[!t]
\includegraphics[width=\linewidth]{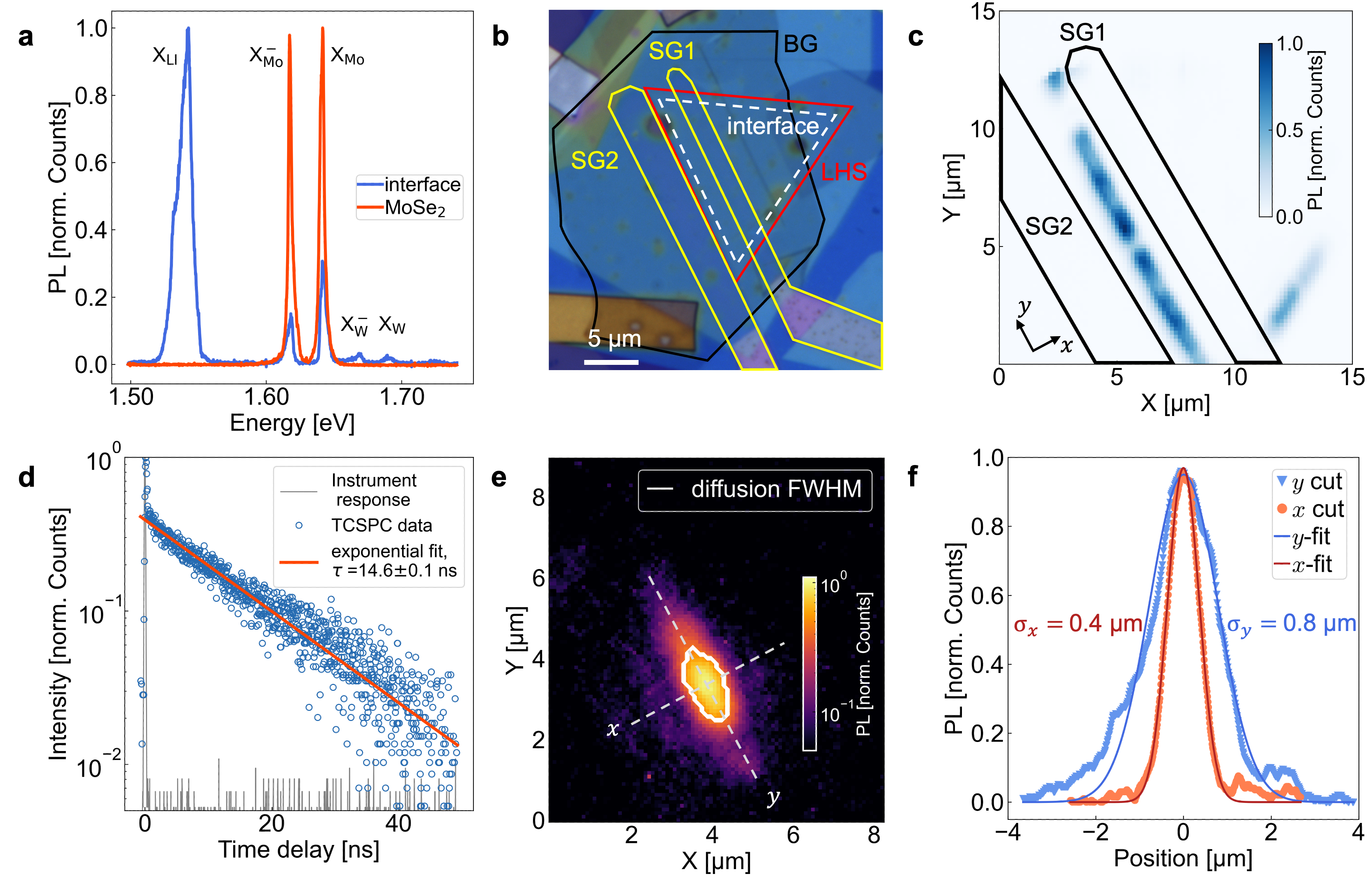}
\caption{\textbf{The interfacial exciton in the lateral heterostructure}. 
\textbf{a}, Normalized PL spectra at 4 K from the 2D bulk MoSe$_2$ (red) and from the MoSe$_2$-WSe$_2$ interface (blue). A new state, $X_{\mathrm{LI}}$, emerges at the interface at $\approx 1.53\,\mathrm{eV}$ ($\sim 100\,\mathrm{meV}$ below the 2D neutral exciton $X_{\mathrm{Mo}}$). 
\textbf{b}, Micrograph of the device: The LHS flake, interface, and gate electrodes (BG, SG) are outlined. 
\textbf{c}, Position-resolved PL integrated over $X_{\mathrm{LI}}$ demonstrates that the state exists exclusively at the 1D interface. 
\textbf{d}, Time-correlated single-photon counting (TCSPC) trace of the $X_{\mathrm{LI}}$ emission. Fitting the long-time decay yields a lifetime of $\tau = 14.6\,\mathrm{ns}$. 
\textbf{e}, Spatial PL diffusion map of the $X_{\mathrm{LI}}$ state obtained by wide-field imaging. The white contour is the Full Width at Half Maximum (FWHM) of the emission spot.
\textbf{f}, Line cuts of the diffusion profile along the short ($x$) and long ($y$) axes. Gaussian fits confirm pronounced anisotropic exciton diffusion, yielding a diffusion length of $L_D \approx 0.7$ µm along $y$ and an estimated diffusion coefficient of $D_y \sim 0.4\,\mathrm{cm}^2\mathrm{s}^{-1}$.}
\label{fig:two}
\end{figure*}

\vspace{-0.3cm}
\subsection*{Excitons at the 1D interface}
We begin by experimentally identifying the interfacial exciton state and establishing its spatial, spectral, and temporal signatures at the 1D junction. In Fig.\,\ref{fig:two}\,\bfA, we present normalized photoluminescence (PL) spectra from the 2D $\text{MoSe}_2$ region (red) and the interface (blue). While the 2D spectrum exhibits the characteristic signatures of the neutral exciton ($X_{\text{Mo}}$) and trion ($X^-_{\text{Mo}}$), measurements at the interface reveal a dominant new resonance at $E_{X_{\text{LI}}} \approx 1.53\,\text{eV}$. This peak is absent in the bulk and lies nearly $100\,\text{meV}$ below the $\text{MoSe}_2$ exciton, placing it well below any known optically active states or defect-bound excitons in the parent monolayers \cite{Srivastava2015, Chakraborty2016}. This significant redshift is consistent with the predicted Type-II band alignment and represents the clear spectral fingerprint of a spatially indirect interfacial state \cite{Urbaszek2023}.

\noindent
\emph{Spatial structure:} To map the spatial distribution of this emission, we perform position-resolved PL imaging across the heterostructure. For reference, a micrograph of the device is shown in Fig.\,\ref{fig:two}\,\bfB, with outlines of the LHS flake (red) and interface (dashed white). As shown in Fig.\,\ref{fig:two}\,\bfC, the $X_{\text{LI}}$ emits exclusively from the 1D junction region, extending continuously along its $\sim 20$ µm length. The emission pattern is convolved with the point spread function of the imaging system, leading to a diffraction limited width. We attribute variations in local intensity to optical absorption by the split-gate electrodes and quenching near the graphene contacts \cite{Lorchat2020}. These spatially resolved measurements unambiguously demonstrate that $X_{\text{LI}}$ is not a local defect but a delocalized 1D state emerging from the crystallographic boundary.

\noindent
\emph{Lifetime:} To probe the temporal dynamics of this state, we employ time-correlated single-photon counting (TCSPC). Fitting the filtered $X_{\text{LI}}$ emission with a single-exponential decay model yields a radiative lifetime of $\tau = 14.6 \pm 0.1\,\text{ns}$ (Fig.\,\ref{fig:two}\,\bfD). This value is three orders of magnitude longer than the lifetimes of intralayer exciton species in the same system ($\sim 1\text{--}100\,\text{ps}$) \cite{Robert2016}. This prolonged lifetime is the hallmark of a charge-transfer exciton state: the spatial separation of the electron and hole wavefunctions across the interface reduces their radiative recombination rate, enhancing their lifetime (see SI for details).

\noindent\emph{Mobility:} We further examine the spatial transport of these 1D excitons under continuous-wave (CW) excitation. The mobility of $X_\mathrm{LI}$ serves as a critical metric to confirm their nature as genuine interfacial quasiparticles. Steady-state PL imaging of the $X_\mathrm{LI}$ (Fig.\,\ref{fig:two}\,\bfE) reveals a markedly anisotropic `cigar-shaped' profile, with exciton spreading occurring primarily in the longitudinal ($y$) direction. Line cuts taken along $x$ and $y$ yield half-widths of (standard deviations, see SI) $\sigma_{y} = 0.8 \pm 0.04$ µm and $\sigma_{x} = 0.4 \pm 0.04$ µm, respectively. While the transverse width $\sigma_{x}$ remains near the diffraction limit of our system ($\sigma_\mathrm{PSF} = 0.35$ µm), the longitudinal spread $\sigma_{y}$ shows a macroscopic signature of 1D motion.

We determine the longitudinal diffusion length by deconvolving the system's point-spread function \cite{Kumar2014,Sun2022,Wietek2024}: $L_D = \sqrt{\sigma_{y}^2 - \sigma_\mathrm{PSF}^2} = 0.72 \pm 0.06$ µm. Using the measured exciton lifetime ($\tau = 15\,\text{ns}$), we estimate an effective 1D diffusion coefficient $D_y = L_D^2/\tau \approx 0.4\,\text{cm}^2/\text{s}$. In the low-excitation power regime, where $D_y$ is independent of pump power, we extract a corresponding exciton mobility $\mu_X = eD_y/k_\mathrm{B}T \approx 1,200\,\text{cm}^2/\text{Vs}$ at $T = 4\,\text{K}$. 

We emphasize that steady-state imaging provides a time-integrated assessment of transport, and further time-resolved measurements would be required to fully map the diffusion dynamics. Nonetheless, such efficient transport properties are a key indicator of the high-quality interfacial 1D channel \cite{Crochet2012,Li2023}. In contrast to 2D monolayers, where ubiquitous small-angle scattering limits mobility, 1D kinematics restrict momentum relaxation to discrete backscattering events. This effectively filters out the long-range potential fluctuations common in van der Waals heterostructures \cite{Wu2016,Moore2020}, suggesting that the seamless lateral interface provides a uniquely protected environment for excitonic transport. Taken together, these signatures provide compelling evidence for long-lived, mobile, interfacial excitonic quasiparticles. \vspace{-0.2cm}

\begin{figure*}[!t]
\includegraphics[width=\linewidth]{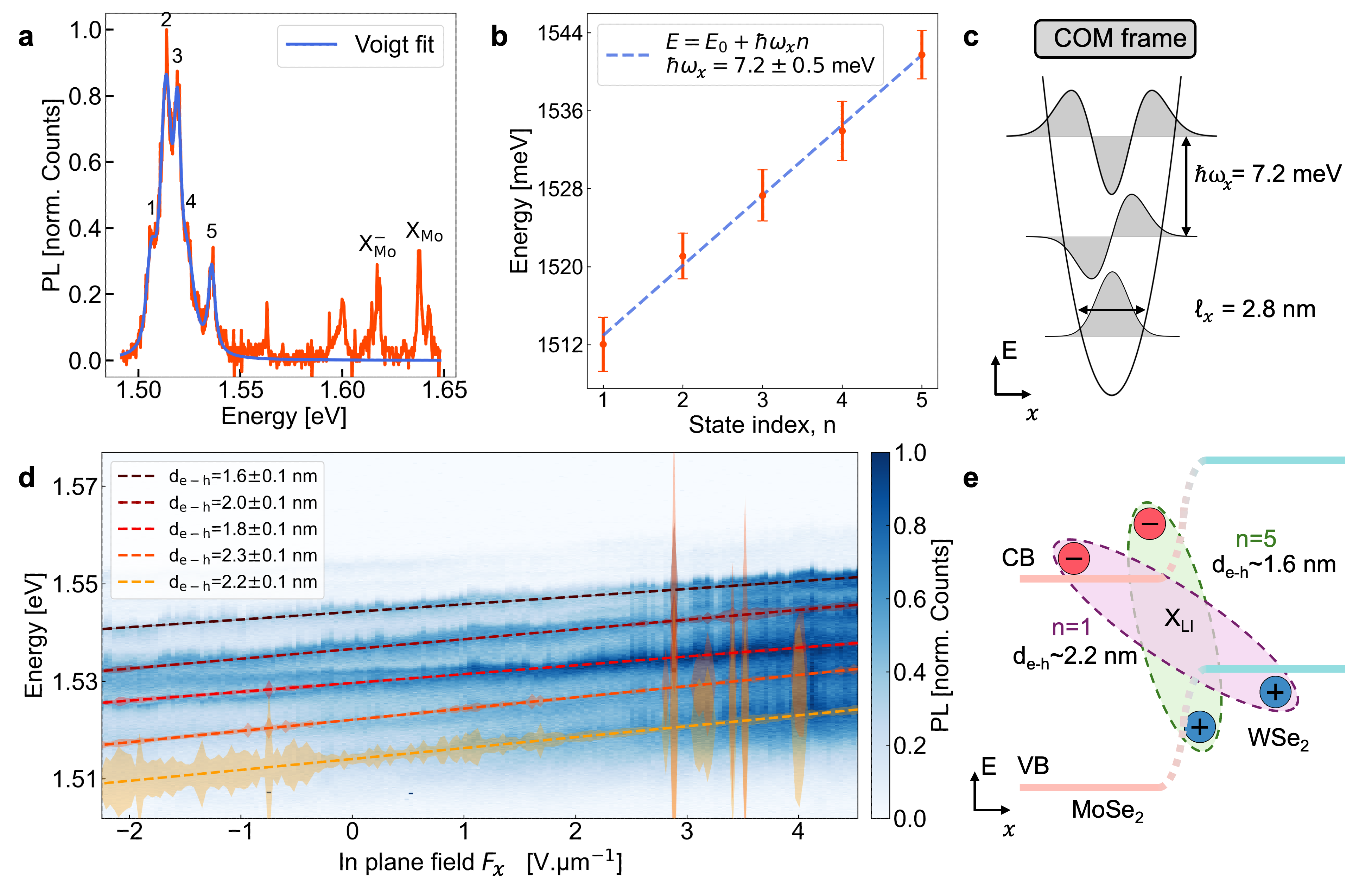}
\caption{\textbf{Quantized 1D Motion and Permanent Dipole Moment of the Interfacial Exciton}. 
\textbf{a}, Low-power PL spectra showing the $X_{\mathrm{LI}}$ state resolved into a ladder of narrow, discrete quantum peaks -- an unambiguous signature of motional quantization. 
\textbf{b}, Energy of the discrete states ($E_n$) versus state index ($n$). A linear fit (blue dashed line), $E_n = E_0 + \hbar\omega_x n$, yields a harmonic level spacing of $\hbar\omega_x = 7.2 \pm 0.5\,\mathrm{meV}$, implying a transverse confinement length of $\ell_x \approx 2.8\,\mathrm{nm}$.  
\textbf{c}, Illustration of the exciton COM wavefunctions confined in a parabolic potential imposed by the band offset. 
\textbf{d}, PL spectra as a function of the in-plane electric field ($\vec{F}_x$) applied perpendicular to the interface via the split gates. The clear, systematic linear shift (Stark shift) of the peak energies is the definitive signature of a permanent in-plane dipole moment. Linear fits (dashed lines) yield the dipole length $d_\mathrm{e\text{-}h}^{(n)}$.
\textbf{e}, Schematic illustrating the reduction of extracted dipole lengths with increasing energy, from $d_\mathrm{e\text{-}h} \approx 2.2\,\mathrm{nm}$ for the lowest state to $\approx 1.6\,\mathrm{nm}$ for higher COM states. }
\label{fig:three}
\end{figure*}

\subsection*{1D quantum confinement and dipole moment}
\vspace{-0.2cm}
We now investigate the properties of the interfacial exciton wavefunction, both in the center-of-mass (COM) and the internal relative degrees of freedom. A closer examination of the $X_{\mathrm{LI}}$ lineshape in Fig.\ref{fig:two}\,\bfA\, suggests a fine structure. By reducing the excitation power ($\lesssim 100\,\mathrm{nW}$), thus minimizing power-broadening, we observe that the broad emission resolves into a ladder of narrow, discrete peaks with linewidths $\Gamma \approx 1\text{--}3\,\mathrm{meV}$, as shown in Fig.\,\ref{fig:three}\,\bfA. While individual peak energies vary slightly across the sample, the characteristic ladder structure and energy splittings are highly reproducible along the entire junction. This spatial robustness, combined with the diffusion signatures, rules out an origin in random, isolated defect states.

The appearance of this discrete ladder constitutes a definitive spectroscopic signature of quantized center-of-mass (COM) motion \cite{Thureja2022, Chervy2024}, establishing $X_{\mathrm{LI}}$ as an intrinsically 1D quantum wire state. The confinement arises from the Type-II band alignment: as the exciton moves away from the junction, the carriers encounter steep band-edge barriers. The exciton thus experiences an effective confining potential: \vspace{-0.2cm}
\begin{equation*}
    V_{\mathrm{conf}}(X) = -V_\mathrm{vb}(x_h) + V_\mathrm{cb}(x_e),
\end{equation*}
where $V_\mathrm{vb}$ and $V_\mathrm{cb}$ are the valence and conduction band offsets, $x_h$ and $x_e$ are coordinates of hole and electron, and $X = (m_ex_e + m_hx_h)/(m_e+m_h)$ denotes the exciton's COM coordinate. Consequently, the potential inherits the full band-offset energy scale ($\approx 0.1 - 0.2\,\mathrm{eV}$) and the atomic-scale width of the interface, creating a remarkably deep and tight 1D trap for the bound state.

In Fig.\,\ref{fig:three}\,\bfB, we plot the center energies of these states extracted from Voigt fits averaged across seven different spatial positions. The energies follow a linear trend, $E_n = E_0 + \hbar\omega_x n$, for the lowest states, yielding a harmonic level spacing of $\hbar\omega_x = 7.2 \pm 0.5\,\mathrm{meV}$. This corresponds to a nanoscopic transverse confinement length of $\ell_x = \sqrt{\hbar/(m_\mathrm{X}\omega_x)} = 2.8 \pm 0.1\,\mathrm{nm}$ (using $m_\mathrm{X} = 1.3\,m_0$) as shown in Fig.\,\ref{fig:three}\,\bfC. Although the underlying band offset step is atomically sharp, the effective COM potential experienced by the exciton is smoothed over the spatial extent of the bound electron–hole pair, yielding an approximately harmonic confinement for the lowest-energy states.  When compared to the micron-scale longitudinal diffusion, this yields an extraordinary aspect ratio exceeding 500:1, underscoring the 1D character of the $X_{\mathrm{LI}}$ state.

Next, we probe the in-plane dipole moment, $\vec{p} = e\vec{d}_\mathrm{e\text{-}h}$, of the $X_{\mathrm{LI}}$ exciton. By applying voltage ($V_\mathrm{SG1}$, $V_\mathrm{SG2}$) on split-gate electrodes (Fig.\,\ref{fig:one}\,\bfE), we apply a controlled in-plane electric field $F_x = (V_\mathrm{SG1} - V_\mathrm{SG2})/D_\mathrm{SG}$ perpendicular to the interface, where $D_\mathrm{SG} = 2.7\,\mu$m is the split-gate separation. We ensure that these measurements are always performed at neutrality (SI). 
For a dipole with a fixed orientation, an anti-aligned field produces a blue shift, while an aligned field induces a red shift. The PL spectra as a function of $F_x$ (Fig.\,\ref{fig:three}\,\bfD) reveal a systematic linear Stark shift—the definitive signature of a permanent dipole moment. Linear fits, $E_n = E_0 + e \, d_\mathrm{e\text{-}h}(n) F_x$, allow us to precisely extract the permanent dipole length $d_\mathrm{e\text{-}h}(n)$ for each COM state. These fits yield dipole lengths of $d_\mathrm{e\text{-}h} \sim 2.2\,\mathrm{nm}$ for the lowest state ($n=1$), decreasing systematically to $\sim 1.6\,\mathrm{nm}$ for higher state ($n=5$), as summarized in Fig.\,\ref{fig:three}\,\bfE. This dipole length is exceptionally large for a bound exciton in TMDs, implying strong dipole–dipole interactions that scale as $U\propto d_\mathrm{e\text{-}h}^2$.

Critically, since the dipole length is comparable to the transverse confinement scale ($\ell_x \approx d_\mathrm{e\text{-}h}$), our observations place the interfacial exciton in the strong-confinement regime. In this limit, the conventional separation of COM and relative motion breaks down; the quasiparticle must be treated as a composite object whose internal wavefunction is ``reshaped" by the external potential. The systematic variation of dipole length across the $n$-states is a direct manifestation of this coupled COM-relative dynamics \cite{Szafran2022}. This regime stands in sharp contrast to traditional 1D platforms—such as GaAs T-wires or gate-defined TMD wires—where $\ell_x \gg a_B$, and the internal structure remains effectively ``frozen." 

\begin{figure*}[!t]
\includegraphics[width=\linewidth]{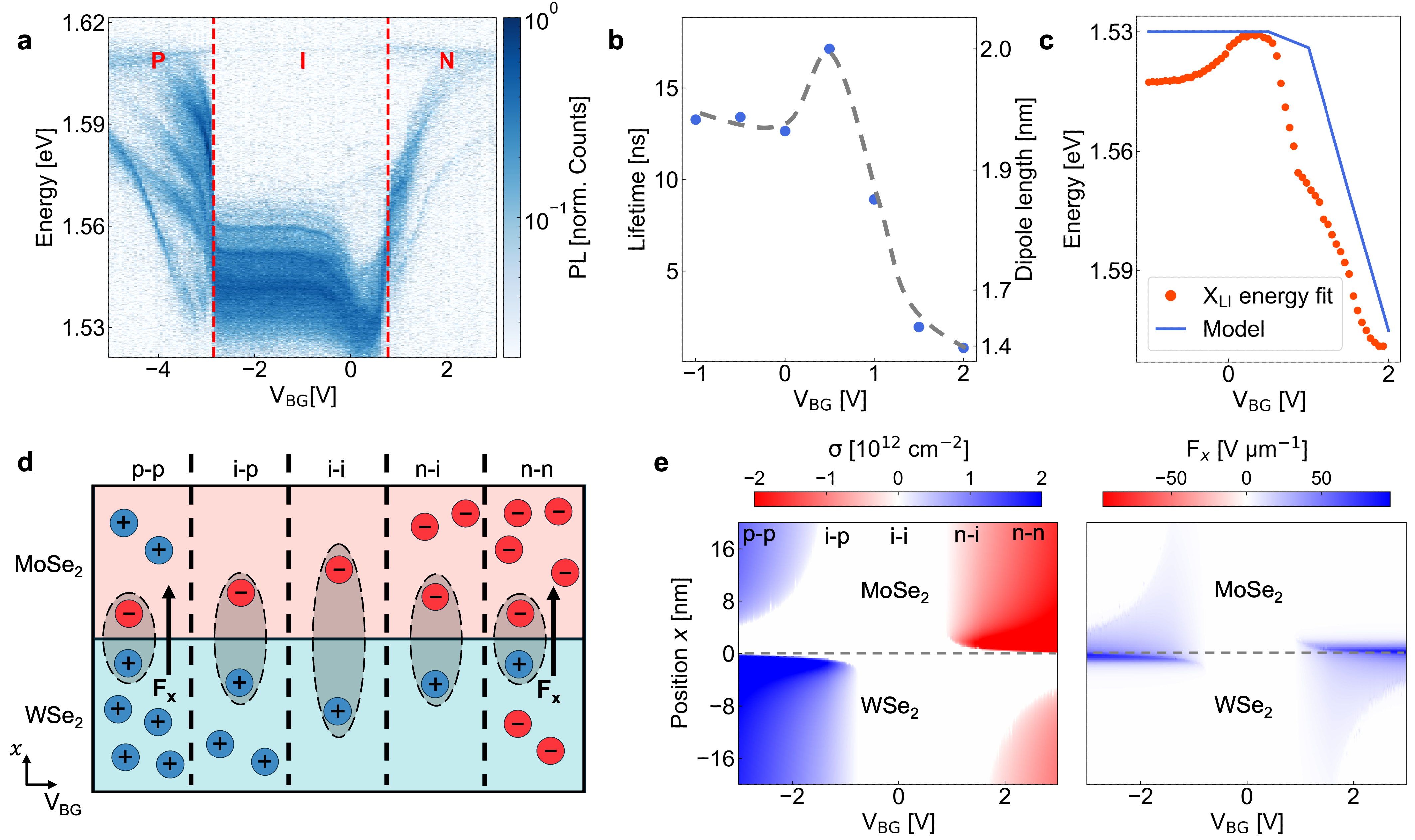}
\caption{\textbf{Dynamic Control over Dipole Moment and Lifetime via Electrostatic Doping}. 
\textbf{a}, PL spectra as a function of the bottom gate voltage ($\vbg$) show a large, systematic blue shift ($\sim 70\,\mathrm{meV}$) of the $X_{\mathrm{LI}}$ state upon charge doping. 
\textbf{b}, Measured $X_{\mathrm{LI}}$ lifetime ($\tau$, blue circles, left axis) on the n-doping side reduces by more than an order of magnitude from $15\,\mathrm{ns}$ at neutrality to $800\,\mathrm{ps}$ at $V_{\mathrm{BG}}=2\,\mathrm{V}$. The corresponding dipole length ($d_\mathrm{e\text{-}h}$) extracted from $\tau$ (right axis) demonstrates dynamic tunability from $2\,\mathrm{nm}$ to $1.4\,\mathrm{nm}$. 
\textbf{c}, Extracted $X_{\mathrm{LI}}$ center energy (orange points) as a function of $\vbg$. The blue curve represents the computed energy based on the DC Stark shift model, using the dipole lengths extracted independently from the lifetime data in (\bfB) and simulated electric fields (\bfE). The excellent qualitative agreement confirms the model linking lifetime, dipole collapse, and spectral energy. 
\textbf{d}, Schematic illustrating the Type-II alignment under doping: charge accumulation generates an in-plane electric field ($F_{\mathrm{x}}$) across the interface that opposes the permanent dipole moment ($\vec{p}$) of the exciton. 
\textbf{e}, Finite-element simulations confirming the concentration of strong in-plane electric fields near the interface upon doping, which drives the dipole collapse.}

\label{fig:four}
\end{figure*}

\subsection*{Dipole length and lifetime tuning}

Having established the fundamental properties of the interfacial exciton, we now examine its response to electrostatic doping and demonstrate the tunability of its dipolar nature and radiative lifetime over an exceptionally wide range.

In Fig.\,\ref{fig:four}\,\bfA, we present the $X_{\mathrm{LI}}$ photoluminescence (PL) spectra as a function of back-gate voltage ($V_{\mathrm{BG}}$). At charge neutrality ($V_{\mathrm{BG}} \approx 0\,\text{V}$), the spectrum exhibits the discrete ladder of COM states. Upon introducing either electron ($V_{\mathrm{BG}} > 0.5\,\text{V}$) or hole ($V_{\mathrm{BG}} < -2\,\text{V}$) doping, the $X_{\mathrm{LI}}$ state undergoes a sharp, continuous blue shift of $\Delta E \approx 70\,\text{meV}$ within a narrow voltage window ($\sim 1\,\text{V}$). This shift eventually plateaus at higher voltages, approaching the 2D trion energy of monolayer \Mo, as global charge accumulation begins to screen the interface band offset.

To understand the origin of this shift, we measure the $X_{\mathrm{LI}}$ lifetime using TCSPC (Fig.\,\ref{fig:four}\,\bfB) as a function of $\vbg$. We observe a dramatic reduction in the lifetime by more than an order-of-magnitude, dropping from $\tau \approx 15\,\text{ns}$ at neutrality to $\tau \approx  800\,\text{ps}$ at $V_{\mathrm{BG}} = 2\,\text{V}$. Crucially, this lifetime collapse occurs in synchrony with the observed blue shift (Fig.\,\ref{fig:four}\,\bfC), suggesting that both observables probe a common underlying reconfiguration of the exciton’s internal structure.

We propose that this behavior is the direct result of a gate-controlled ``dipole collapse." This effect can be viewed through two complementary yet equivalent lenses: first, gate-induced doping generates a concentrated in-plane electric field $F_x$ that is anti-aligned to the dipole regardless of the doping configuration (see Fig.\,\ref{fig:four}\,\bfD). This field, confirmed by electrostatic simulations, shown in Fig.\,\ref{fig:four}\,\bfE, exerts a compressive force on the dipolar exciton. Alternatively, the onset of doping screens and renormalizes the intrinsic band-offset, effectively flattening the Type-II potential. This screening reduces the electrostatic barrier that partitions the carriers, allowing the electron and hole wavefunctions to penetrate more deeply into the adjacent regions, thereby increasing their spatial overlap.

To quantify this structural change, we utilize a tunneling model \cite{WangYao2018} where the electron and hole wavefunctions are treated as quasi-bound states with evanescent tails: $\psi_{e,h} \sim \exp(\frac{-|x \pm d_\mathrm{e\text{-}h}/2|}{b})$, where $b$ is a tunneling length (see SI for further details). The electron-hole wavefunction overlap relates directly to the dipole length, according to $\mathcal{O}_{eh} = \int \psi_e^*\psi_h dx \propto \exp(\frac{-d_\mathrm{e\text{-}h}}{b})$.  Hence, enhancement of the overlap reduces radiative lifetime ($\tau \propto \mathcal{O}_\mathrm{e\text{-}h}^{-2}$), through the relation: $\tau = \tau_0 \exp\left(\frac{2d_\mathrm{e\text{-}h}}{b}\right)$, where $\tau_0 \approx 1\text{ps}$ is the known homogeneously broadened radiative lifetime of a 1s exciton with maximal overlap ($d_\mathrm{e\text{-}h} = 0$)\cite{Robert2016}. By anchoring this model to our measured reference values of dipole length $\tilde{d}_\mathrm{e\text{-}h} = 2\,\text{nm}$ and lifetime $\tilde{\tau} = 15\,\text{ns}$ at neutrality, we derive a zero-parameter relationship to extract the $\vbg$-dependent dipole length: \vspace{-0.2cm}
\begin{equation}
    d_\mathrm{e\text{-}h}(\vbg) = \tilde{d}_\mathrm{e\text{-}h} \frac{\ln(\tau(\vbg)/\tau_0)}{\ln(\tilde{\tau}/\tau_0)}.
\end{equation}
For simplicity, this model assumes a constant tunneling length $b$. While $b$ may physically vary as the gate-dependent potential renormalization modifies the barrier height, we find that the modulation of the dipole length $d_\mathrm{e\text{-}h}$ is the primary driver of the observed dynamics. The extracted dipole lengths (Fig.\,\ref{fig:four}\,\bfB, right axis) reveal a dynamic tuning of the exciton's internal size from $2\,\text{nm}$ down to $\sim 1.4\,\text{nm}$. To validate this model, we calculate the energy shift, $\Delta E(\vbg) = -e \vec{d}_\mathrm{e\text{-}h}(\vbg) \cdot \vec{F}_{x,\text{max}}(\vbg)$, using only these independently extracted dipole lengths and simulated maximum field amplitude at the interface (Fig.\,\ref{fig:four}\,\bfE). The resulting curve (Fig.\,\ref{fig:four}\,\bfC, blue line) shows remarkable consistency with the experimental data.

While electrostatic doping can influence non-radiative decay, such channels would not typically produce a sharp lifetime collapse in synchrony with a monotonic blue shift. The fact that a structural change inferred solely from temporal dynamics predicts the order of magnitude spectral shift provides definitive evidence for the dipole-collapse mechanism. This result establishes the lateral interface as a unique platform where the internal quantum structure of an exciton can be continuously engineered in-situ. \vspace{-0.3cm}

\subsection*{Discussion}
Our findings establish the lateral $\text{MoSe}_2/\text{WSe}_2$ interface as an intrinsic 1D quantum wire, where quasiparticles emerge via crystallographic control and electron-hole interactions rather than top-down lithography. 

The lateral architecture enables in-situ electrostatic tuning of the exciton’s internal dipole and radiative rate, in contrast to vertical heterostructures where interlayer exciton properties are fixed by layer separation. 

Beyond its structural tunability, the interface represents a promising building block for excitonic technologies. The high mobilities ($\sim 1200\,\text{cm}^2/\text{Vs}$) suggest that interfacial excitons are protected by suppressed momentum relaxation due to reduced dimensionality. Such efficient transport positions these atomically sharp channels as low-loss interconnects and high-performance transistors for future excitonic circuitry.

The 1D localization of $X_{\mathrm{LI}}$ further suggests a role as a non-invasive, local probe of electronic topological order. By residing strictly at the junction, the interfacial exciton serves as a nanometer-scale sensor for edge states and electronic fluctuations localized at the boundary—physics that is typically inaccessible to bulk optical measurements.

Finally, the large permanent dipole moment and stable 1D confinement provide an ideal arena for many-body physics. The strong repulsive interactions and independent control over confinement and density offer a unique ``knob" to realize strongly correlated phases, such as a Tonks-Girardeau gas of dipolar bosons, moving well beyond the dilute-gas limit. In summary, this lateral junction provides a unique platform for the active engineering of 1D excitonic devices and the exploration of novel quantum phases in reduced dimensions.

\newpage
\section*{Methods}
\subsection*{Device fabrication and experimental setup}
Samples are stacked under an inert Ar atmosphere using a dry stacking procedure with PDMS droplets covered by a PC film. The LHS flakes are grown on SiO$_2$ substrates and are picked up, alongside the other graphene and encapsulating flakes, by a top hBN ($\sim25 $nm thick). Graphene and hBN flakes are obtained through manual bulk exfoliation. We contact the samples using prepatterned gold electrodes on Si/SiO$_2$ substrates. After the stack is assembled, 13\,nm-thick gold split gates are deposited on top using e-beam lithography, e-beam evaporation of gold and bilayer liftoff. \\

All measurements are conducted at a temperature of $4\,$K, in a cryogenic confocal microscopy setup that also provides electrical access to the sample. Supplementary Information Fig.\,\textbf{1} presents a schematic overview of the setup.

\subsection*{Lifetime measurements}
The lifetime of interface excitons is measured using Time Correlated Single Photon Counting (TCSPC) method. An excitation laser pulse (pulse duration $150\,$fs, repetition rate $80\,$MHz, average power 100nW , central wavelength 730nm), is focused on the LHS device. The PL emission from the interfacial exciton is spectrally filtered and detected on a superconducting nanowire single photon detector (SNSPD, \emph{Single Quantum}), with a temporal jitter of $\sim 15\,$ps. 

In parallel, a fraction of the excitation laser is strongly attenuated and detected by a second SNSPD detector, providing a reference signal that defines the zero-time delay for PL decay. The electronic output signals from both SNSPDs are recorded in stop-start histogram mode using a time-tagger (\emph{Swabian Instruments}), yielding the TCSPC curves shown in Fig\,\ref{fig:two} of the main text.

The TCSPC data are fit with a single exponential decay model, $I(t) = I_0\exp(-t/\tau) + \delta$, where $\delta$ accounts for the background noise floor. At short time delays, the measured decay is influenced by the instrument response function (IRF). The IRF is independently characterized by removing the spectral filters from the signal collection path, allowing the excitation pulses to be detected by both SNSPDs. The resulting IRF is shown as the gray trace in Fig.\,\ref{fig:two}\,\bfD.

\subsection*{Electrostatic simulation of TMD heterostructure devices}
Finite-element method (FEM) electrostatic simulations provide quantitative information of the in-plane fields and charge densities at a Type-II TMDs lateral interface. These computations are performed using the Electrostatics package in COMSOL. For all our simulations, we assume temperature T = 0 K.\\
Each TMDs are modeled using the Thomas-Fermi approximation as a single sheet of charge with density,
\begin{equation*}
    \sigma_{\text{XSe}_2}(x) = \sigma_{n,{\text{XSe}_2}}(x)+\sigma_{p,{\text{XSe}_2}}(x)
\end{equation*}
where X=Mo,W and $\sigma_{n,{\text{XSe}_2}}(x)$ and $\sigma_{p,{\text{XSe}_2}}(x)$ are electron and hole charge densities respectively, which are--in turn--given by,
\begin{equation*}
    \scalebox{0.8}{$
    \begin{aligned} 
        \sigma_{n,{\text{XSe}_2}}(x) &=
        -e \int_{E_{C,{\text{XSe}_2}}(V(x))}^{E_F}{D_{\text{XSe}_2}\,dE},\,E_F>E_{C,{\text{XSe}_2}}(V(x)) \\
        &= -e\,D_{\text{XSe}_2} (E_F-E_{C,_{\text{XSe}_2}}),
    \end{aligned}$}
\end{equation*}
\begin{equation*}
    \scalebox{0.8}{$
    \begin{aligned}
       \sigma_{p,{\text{XSe}_2}}(x) &=
       e \int_{E_F}^{E_{V,{\text{XSe}_2}}(V(x))}{D_{\text{XSe}_2}\,dE},\,E_F<E_{V,{\text{XSe}_2}}(V(x)) \\
       &= e\,D_{\text{XSe}_2} (E_{V,_{\text{XSe}_2}}-E_F).
   \end{aligned}$}
\end{equation*}
Here $D_{\text{XSe}_2}=g_S\,g_V\,m_{\text{XSe}_2}^*/2\pi\hbar^2$ is the 2D density of states for electrons and holes in the TMD semiconductor, where $g_S=1$ is the spin degeneracy and $g_V=2$ is the valley degeneracy. $E_{C,{\text{XSe}_2}}$ and $E_{V,{\text{XSe}_2}}$ are the conduction and valence band edge energies, respectively, which depend on the local electrostatic potential $V(x)$ and on the Type-II band alignment at the ${\text{MoSe}_2}$-${\text{WSe}_2}$ interface. $E_F$ is the Fermi level determined by the alignment of the contact work function with respect to the band edges.\\
The LHS sheet of charge is encapsulated by two 25-nm-thick hBN slabs and contacted by ohmic electrodes. The material parameters assumed for the simulations are given in the SI.
Finally, the bottom gate and split gates are defined in the simulation as fixed-potential electrodes, and voltages are applied to obtain the spatial charge density and electric field distributions across the LHS. The simulation results in Fig.\,\ref{fig:four}\,\bfE\, of the main text are obtained by varying the bottom gate voltage $\vbg$ to dope across the ${\text{MoSe}_2}$-${\text{WSe}_2}$ interface. Separate simulations with only the split gates verify that the electric field at the junction is purely in-plane.

\subsection*{Spectral analysis procedure}
The PL intensity as a function of energy, \( I(E) \), is modeled as the sum of five Voigt profiles:
\[
I(E) = \sum_{i=1}^{5} A_i \cdot V(E; E_{0,i}, \sigma_i, \gamma_i)
\]
where the Voigt profile \( V(E; E_{0,i}, \sigma_i, \gamma_i) \) is defined as the convolution of a Gaussian and a Lorentzian.
Supplementary Information Fig.\,\textbf{11} gives a visualization of the individual Voigt profiles summing up to the spectral profile shown in Fig.\,\ref{fig:three}\,\bfE\, of the main text. The average Lorentzian linewidth extracted from these peaks is \( \bar\gamma = 2\,\text{meV}\). 

\section*{References}
  \addcontentsline{toc}{section}{References}

{\small
\printbibliography[heading=none]

@article{Szafran2022,
  title = {Exciton localization on $p\ensuremath{-}i\ensuremath{-}n$ junctions in two-dimensional crystals},
  author = {Szafran, Bart\l{}omiej},
  journal = {Phys. Rev. B},
  volume = {106},
  issue = {8},
  pages = {085305},
  numpages = {8},
  year = {2022},
  month = {08},
  publisher = {American Physical Society},
}

@article{Wang2018,
  title = {Colloquium: Excitons in atomically thin transition metal dichalcogenides},
  author = {Wang, Gang and Chernikov, Alexey and Glazov, Mikhail M. and Heinz, Tony F. and Marie, Xavier and Amand, Thierry and Urbaszek, Bernhard},
  journal = {Rev. Mod. Phys.},
  volume = {90},
  issue = {2},
  pages = {021001},
  numpages = {25},
  year = {2018},
  month = {04},
  publisher = {American Physical Society},
}

@article{Sahoo2018,
author={Sahoo, Prasana K.
and Memaran, Shahriar
and Xin, Yan
and Balicas, Luis
and Guti{\'e}rrez, Humberto R.},
title={One-pot growth of two-dimensional lateral heterostructures via sequential edge-epitaxy},
journal={Nature},
year={2018},
volume={553},
number={7686},
pages={63-67},}

@article{WangYao2018,
title = {Interface excitons at lateral heterojunctions in monolayer semiconductors},
author = {Lau, Ka Wai and Calvin and Gong, Zhirui and Yu, Hongyi and Yao, Wang},
journal = {Phys. Rev. B},
volume = {98},
issue = {11},
pages = {115427},
numpages = {15},
year = {2018},}

@article{Urbaszek2023,
author={Rosati, Roberto
and Paradisanos, Ioannis
and Huang, Libai
and Gan, Ziyang
and George, Antony
and Watanabe, Kenji
and Taniguchi, Takashi
and Lombez, Laurent
and Renucci, Pierre
and Turchanin, Andrey
and Urbaszek, Bernhard
and Malic, Ermin},
title={Interface engineering of charge-transfer excitons in 2D lateral heterostructures},
journal={Nature Communications},
year={2023},
volume={14},
number={1},
pages={2438}}

@article{Yuan2023,
author = {Yuan, Long and Zheng, Biyuan and Zhao, Qiuchen and Kempt, Roman and Brumme, Thomas and Kuc, Agnieszka B. and Ma, Chao and Deng, Shibin and Pan, Anlian and Huang, Libai},
title = {Strong Dipolar Repulsion of One-Dimensional Interfacial Excitons in Monolayer Lateral Heterojunctions},
journal = {ACS Nano},
volume = {17},
number = {16},
pages = {15379-15387},
year = {2023}}

@article{DUrnevSmirnov2025,
title = {Intervalley mixing of interface excitons at lateral heterojunctions},
author = {Durnev, M. V. and Smirnov, D. S.},
journal = {Phys. Rev. B},
volume = {111},
issue = {20},
pages = {205403},
numpages = {11},
year = {2025},
publisher = {American Physical Society}}

@article{Sahoo2024,
author={Kundu, Baisali
and Mondal, Priyanka
and Tebbe, David
and Hasan, Md. Nur
and Chakraborty, Suman Kumar
and Metzelaars, Marvin
and K{\"o}gerler, Paul
and Karmakar, Debjani
and Pradhan, Gopal K.
and Stampfer, Christoph
and Beschoten, Bernd
and Waldecker, Lutz
and Sahoo, Prasana Kumar},
title={Electrically Controlled Excitons, Charge Transfer Induced Trions, and Narrowband Emitters in MoSe2--WSe2 Lateral Heterostructure},
journal={Nano Letters},
year={2024},
publisher={American Chemical Society},
volume={24},
number={46},
pages={14615-14624}}

@article{Wilson2017,
author = {Neil R. Wilson  and Paul V. Nguyen  and Kyle Seyler  and Pasqual Rivera  and Alexander J. Marsden  and Zachary P. L. Laker  and Gabriel C. Constantinescu  and Viktor Kandyba  and Alexei Barinov  and Nicholas D. M. Hine  and Xiaodong Xu  and David H. Cobden },
title = {Determination of band offsets, hybridization, and exciton binding in 2D semiconductor heterostructures},
journal = {Science Advances},
volume = {3},
number = {2},
pages = {e1601832},
year = {2017}}

@article{Robertson2016,
author = {Guo, Yuzheng and Robertson, John},
title = {Band engineering in transition metal dichalcogenides: Stacked versus lateral heterostructures},
journal = {Applied Physics Letters},
volume = {108},
number = {23},
pages = {233104},
year = {2016}}

@article{Degani1990,
title = {Exciton binding energy in type-II heterojunctions},
author = {Degani, Marcos H. and Farias, Gil A.},
journal = {Phys. Rev. B},
volume = {42},
issue = {18},
pages = {11701--11707},
numpages = {0},
year = {1990},
publisher = {American Physical Society}}

@article{Richard2022,
title = {Excitonic Tonks-Girardeau and charge density wave phases in monolayer semiconductors},
author = {O\l{}dziejewski, Rafa\l{} and Chiocchetta, Alessio and Kn\"orzer, Johannes and Schmidt, Richard},
journal = {Phys. Rev. B},
volume = {106},
issue = {8},
pages = {L081412},
numpages = {6},
year = {2022},
publisher = {American Physical Society}}

@Article{Thureja2022,
author={Thureja, Deepankur
and Imamoglu, Atac
and Smole{\'{n}}ski, Tomasz
and Amelio, Ivan
and Popert, Alexander
and Chervy, Thibault
and Lu, Xiaobo
and Liu, Song
and Barmak, Katayun
and Watanabe, Kenji
and Taniguchi, Takashi
and Norris, David J.
and Kroner, Martin
and Murthy, Puneet A.},
title={Electrically tunable quantum confinement of neutral excitons},
journal={Nature},
year={2022},
volume={606},
number={7913},
pages={298-304}}

@article{
Chervy2024,
author = {Jenny Hu  and Etienne Lorchat  and Xueqi Chen  and Kenji Watanabe  and Takashi Taniguchi  and Tony F. Heinz  and Puneet A. Murthy  and Thibault Chervy },
title = {Quantum control of exciton wave functions in 2D semiconductors},
journal = {Science Advances},
volume = {10},
number = {12},
pages = {eadk6369},
year = {2024}}

@Article{Jiang2021,
author={Jiang, Ying
and Chen, Shula
and Zheng, Weihao
and Zheng, Biyuan
and Pan, Anlian},
title={Interlayer exciton formation, relaxation, and transport in TMD van der Waals heterostructures},
journal={Light: Science {\&} Applications},
year={2021},
volume={10},
number={1},
pages={72}}

@Article{Sun2022,
author={Sun, Zhe
and Ciarrocchi, Alberto
and Tagarelli, Fedele
and Gonzalez Marin, Juan Francisco
and Watanabe, Kenji
and Taniguchi, Takashi
and Kis, Andras},
title={Excitonic transport driven by repulsive dipolar interaction in a van der Waals heterostructure},
journal={Nature Photonics},
year={2022},
volume={16},
number={1},
pages={79-85}}

@Article{Rivera2015,
author={Rivera, Pasqual
and Schaibley, John R.
and Jones, Aaron M.
and Ross, Jason S.
and Wu, Sanfeng
and Aivazian, Grant
and Klement, Philip
and Seyler, Kyle
and Clark, Genevieve
and Ghimire, Nirmal J.
and Yan, Jiaqiang
and Mandrus, D. G.
and Yao, Wang
and Xu, Xiaodong},
title={Observation of long-lived interlayer excitons in monolayer MoSe2--WSe2 heterostructures},
journal={Nature Communications},
year={2015},
volume={6},
number={1},
pages={6242}}

@Article{Duan2014,
author={Duan, Xidong
and Wang, Chen
and Shaw, Jonathan C.
and Cheng, Rui
and Chen, Yu
and Li, Honglai
and Wu, Xueping
and Tang, Ying
and Zhang, Qinling
and Pan, Anlian
and Jiang, Jianhui
and Yu, Ruqing
and Huang, Yu
and Duan, Xiangfeng},
title={Lateral epitaxial growth of two-dimensional layered semiconductor heterojunctions},
journal={Nature Nanotechnology},
year={2014},
volume={9},
number={12},
pages={1024-1030}}

@Article{Huang2014,
author={Huang, Chunming
and Wu, Sanfeng
and Sanchez, Ana M.
and Peters, Jonathan J. P.
and Beanland, Richard
and Ross, Jason S.
and Rivera, Pasqual
and Yao, Wang
and Cobden, David H.
and Xu, Xiaodong},
title={Lateral heterojunctions within monolayer MoSe2--WSe2 semiconductors},
journal={Nature Materials},
year={2014},
volume={13},
number={12},
pages={1096-1101}}

@article{Li2015,
author = {Ming-Yang Li  and Yumeng Shi  and Chia-Chin Cheng  and Li-Syuan Lu  and Yung-Chang Lin  and Hao-Lin Tang  and Meng-Lin Tsai  and Chih-Wei Chu  and Kung-Hwa Wei  and Jr-Hau He  and Wen-Hao Chang  and Kazu Suenaga  and Lain-Jong Li },
title = {Epitaxial growth of a monolayer WSe2-MoS2 lateral p-n junction with an atomically sharp interface},
journal = {Science},
volume = {349},
number = {6247},
pages = {524-528},
year = {2015}}

@Article{Beret2022,
author={Beret, Dorian
and Paradisanos, Ioannis
and Lamsaadi, Hassan
and Gan, Ziyang
and Najafidehaghani, Emad
and George, Antony
and Lehnert, Tibor
and Biskupek, Johannes
and Kaiser, Ute
and Shree, Shivangi
and Estrada-Real, Ana
and Lagarde, Delphine
and Marie, Xavier
and Renucci, Pierre
and Watanabe, Kenji
and Taniguchi, Takashi
and Weber, S{\'e}bastien
and Paillard, Vincent
and Lombez, Laurent
and Poumirol, Jean-Marie
and Turchanin, Andrey
and Urbaszek, Bernhard},
title={Exciton spectroscopy and unidirectional transport in MoSe2-WSe2 lateral heterostructures encapsulated in hexagonal boron nitride},
journal={npj 2D Materials and Applications},
year={2022},
volume={6},
number={1},
pages={84}}

@article{Kang2013,
author = {Kang, Jun and Tongay, Sefaattin and Zhou, Jian and Li, Jingbo and Wu, Junqiao},
title = {Band offsets and heterostructures of two-dimensional semiconductors},
journal = {Applied Physics Letters},
volume = {102},
number = {1},
pages = {012111},
year = {2013}}

@article{Chu2018,
author = {Chu, Yu-Hsun and Wang, Li-Hong and Lee, Shin-Ye and Chen, Hou-Ju and Yang, Po-Ya and Butler, Christopher J. and Lu, Li-Syuan and Yeh, Han and Chang, Wen-Hao and Lin, Minn-Tsong},
title = {Atomic scale depletion region at one dimensional MoSe2-WSe2 heterointerface},
journal = {Applied Physics Letters},
volume = {113},
number = {24},
pages = {241601},
year = {2018}}

@article{Srivastava2015,
author={Srivastava, Ajit
and Sidler, Meinrad
and Allain, Adrien V.
and Lembke, Dominik S.
and Kis, Andras
and Imamo{\u{g}}lu, A.},
title={Optically active quantum dots in monolayer WSe2},
journal={Nature Nanotechnology},
year={2015},
month={06},
day={01},
volume={10},
number={6},
pages={491-496},
}

@article{Cazalilla2011,
  title = {One dimensional bosons: From condensed matter systems to ultracold gases},
  author = {Cazalilla, M. A. and Citro, R. and Giamarchi, T. and Orignac, E. and Rigol, M.},
  journal = {Rev. Mod. Phys.},
  volume = {83},
  issue = {4},
  pages = {1405--1466},
  year = {2011},
  month = {12},
  publisher = {American Physical Society}}

@article{Lorchat2020,
author={Lorchat, Etienne
and L{\'o}pez, Luis E. Parra
and Robert, C{\'e}dric
and Lagarde, Delphine
and Froehlicher, Guillaume
and Taniguchi, Takashi
and Watanabe, Kenji
and Marie, Xavier
and Berciaud, St{\'e}phane},
title={Filtering the photoluminescence spectra of atomically thin semiconductors with graphene},
journal={Nature Nanotechnology},
year={2020},
volume={15},
number={4},
pages={283-288},
}

@article{Rosati2025,
author={Rosati, Roberto
and Shradha, Sai
and Picker, Julian
and Turchanin, Andrey
and Urbaszek, Bernhard
and Malic, Ermin},
title={Impact of Charge-Transfer Excitons on Unidirectional Exciton Transport in Lateral TMD Heterostructures},
journal={Nano Letters},
year={2025},
volume={25},
number={29},
pages={11319-11324},
}

@article{Chakraborty2016,
author = {Chitraleema Chakraborty and Kenneth M. Goodfellow and A. Nick Vamivakas},
journal = {Opt. Mater. Express},
number = {6},
pages = {2081--2087},
title = {Localized emission from defects in MoSe2 layers},
volume = {6},
year = {2016}}

@article{Robert2016,
  title = {Exciton radiative lifetime in transition metal dichalcogenide monolayers},
  author = {Robert, C. and Lagarde, D. and Cadiz, F. and Wang, G. and Lassagne, B. and Amand, T. and Balocchi, A. and Renucci, P. and Tongay, S. and Urbaszek, B. and Marie, X.},
  journal = {Phys. Rev. B},
  volume = {93},
  pages = {205423},
  numpages = {10},
  year = {2016}}

@Article{Kumar2014,
author ="Kumar, Nardeep and Cui, Qiannan and Ceballos, Frank and He, Dawei and Wang, Yongsheng and Zhao, Hui",
title  ="Exciton diffusion in monolayer and bulk MoSe2",
journal  ="Nanoscale",
year  ="2014",
volume  ="6",
pages  ="4915-4919"}

@article{Wietek2024,
  title = {Nonlinear and Negative Effective Diffusivity of Interlayer Excitons in Moir\'e-Free Heterobilayers},
  author = {Wietek, Edith and Florian, Matthias and G\"oser, Jonas and Taniguchi, Takashi and Watanabe, Kenji and H\"ogele, Alexander and Glazov, Mikhail M. and Steinhoff, Alexander and Chernikov, Alexey},
  journal = {Phys. Rev. Lett.},
  volume = {132},
  pages = {016202},
  year = {2024}}

@article{Kane1998,
    author = {Kane, B. E. and Facer, G. R. and Dzurak, A. S. and Lumpkin, N. E. and Clark, R. G. and Pfeiffer, L. N. and West, K. W.},
    title = {Quantized conductance in quantum wires with gate-controlled width and electron density},
    journal = {Applied Physics Letters},
    volume = {72},
    number = {26},
    pages = {3506-3508},
    year = {1998}}

@article{lima2023,
  title = {Tuning of exciton type by environmental screening},
  author = {Lima, Igor L. C. and Milo\ifmmode \check{s}\else \v{s}\fi{}evi\ifmmode \acute{c}\else \'{c}\fi{}, M. V. and Peeters, F. M. and Chaves, Andrey},
  journal = {Phys. Rev. B},
  volume = {108},
  pages = {115303},
  year = {2023}}

@article{Wu2016,
author = {Di Wu  and Xiao Li  and Lan Luan  and Xiaoyu Wu  and Wei Li  and Maruthi N. Yogeesh  and Rudresh Ghosh  and Zhaodong Chu  and Deji Akinwande  and Qian Niu  and Keji Lai },
title = {Uncovering edge states and electrical inhomogeneity in MoS2 field-effect transistors},
journal = {Proceedings of the National Academy of Sciences},
volume = {113},
number = {31},
pages = {8583-8588},
year = {2016}}

@Article{Moore2020,
author={Moore, David
and Jo, Kiyoung
and Nguyen, Christine
and Lou, Jun
and Muratore, Christopher
and Jariwala, Deep
and Glavin, Nicholas R.},
title={Uncovering topographically hidden features in 2D MoSe2 with correlated potential and optical nanoprobes},
journal={npj 2D Materials and Applications},
year={2020},
volume={4},
number={1},
pages={44}}

@article{Dirnberger2021,
author = {Florian Dirnberger  and Jonas D. Ziegler  and Paulo E. Faria Junior  and Rezlind Bushati  and Takashi Taniguchi  and Kenji Watanabe  and Jaroslav Fabian  and Dominique Bougeard  and Alexey Chernikov  and Vinod M. Menon },
title = {Quasi-1D exciton channels in strain-engineered 2D materials},
journal = {Science Advances},
volume = {7},
number = {44},
year = {2021}}

@article{Colombo2008,
title = {Ga-assisted catalyst-free growth mechanism of GaAs nanowires by molecular beam epitaxy},
author = {Colombo, C. and Spirkoska, D. and Frimmer, M. and Abstreiter, G. and Fontcuberta i Morral, A.},
journal = {Phys. Rev. B},
volume = {77},
pages = {155326},
year = {2008}}

@Article{Li2025,
author ="Li, Yunfei and Hu, Ziyi and Guo, Qing and Li, Jing and Liu, Shuai and Xie, Xiaoxuan and Zhang, Xu and Kang, Lixing and Li, Qingwen",
title  ="van der Waals one-dimensional atomic crystal heterostructure derived from carbon nanotubes",
journal  ="Chem. Soc. Rev.",
year  ="2025",
volume  ="54",
pages  ="5619-5656"}

@book{Giamarchi2003,
    author = {Giamarchi, Thierry},
    title = {Quantum Physics in One Dimension},
    publisher = {Oxford University Press},
    year = {2003}}

@article{Sakaki1980,
year = {1980},
volume = {19},
number = {12},
pages = {L735},
author = {Sakaki, Hiroyuki},
title = {Scattering Suppression and High-Mobility Effect of Size-Quantized Electrons in Ultrafine Semiconductor Wire Structures},
journal = {Japanese Journal of Applied Physics}}

@article{Wees1988,
  title = {Quantized conductance of point contacts in a two-dimensional electron gas},
  author = {van Wees, B. J. and van Houten, H. and Beenakker, C. W. J. and Williamson, J. G. and Kouwenhoven, L. P. and van der Marel, D. and Foxon, C. T.},
  journal = {Phys. Rev. Lett.},
  volume = {60},
  pages = {848--850},
  year = {1988}}

@article{Crochet2012,
author = {Crochet, Jared J. and Duque, Juan G. and Werner, James H. and Lounis, Brahim and Cognet, Laurent and Doorn, Stephen K.},
title = {Disorder Limited Exciton Transport in Colloidal Single-Wall Carbon Nanotubes},
journal = {Nano Letters},
volume = {12},
number = {10},
pages = {5091-5096},
year = {2012}}

@article{Li2023,
author = {Li, Zidong and Florian, Matthias and Datta, Kanak and Jiang, Zhaohan and Borsch, Markus and Wen, Qiannan and Kira, Mack and Deotare, Parag B.},
title = {Enhanced Exciton Drift Transport through Suppressed Diffusion in One-Dimensional Guides},
journal = {ACS Nano},
volume = {17},
number = {22},
pages = {22410-22417},
year = {2023}}
}

\noindent
\textbf{Acknowledgements.} We thank Chirag Vaswani, Ouri Karni and Tony F. Heinz for insightful discussions; Lukas Novotny, Jonas Ziegler and Shengyu Shan for experimental support with fabrication, and Anna Herrmann and Iñigo Lasheras for assistance with experiments and fabrication. 

\noindent
\textbf{Funding.} This work was supported by Swiss National Science Foundation (SNSF) Starting Grant no. 211448. K.W. and T.T. acknowledge support from the JSPS KAKENHI (grant numbers 19H05790, 20H00354, and 21H05233). P.S. acknowledges the Department of Science and Technology (DST) (Project Code: DST/NM/TUE/QM-1/2019 and National Quantum Mission (NQM) DST/QTC/NQM/QMD/2024/4/(G)), India. \\

\noindent
\textbf{Author contributions.}
E.V. and F.F. carried out the experiments and analyzed the data. F.F. and E.V. fabricated the samples. F.F. performed the electrostatic simulations, with input from P.A.M. and E.V.. S.K.C, B.N., and P.K.S. performed CVD growth of LHS crystals. K.W. and T.T. provided the h-BN crystals. P.A.M., E.V. and F.F. wrote the manuscript. P.A.M., T.C., and P.K.S  supervised the project.

\noindent
\textbf{Competing interests.} The authors declare no competing interests.

\end{document}